\documentclass[]{aastex631}

\makeatletter
\def\edit#1#2{\ifcase#1\or
\texorpdfstring{#2}{#2}%
\or
\ifturnofftwo\unskip\else\texorpdfstring{{\bfseries\itshape#2}}{#2}\fi
\or
\ifturnoffthree\unskip\else\texorpdfstring{{\bfseries\underline{#2}}}{#2}\fi
\fi}
\makeatother

\usepackage{wrapfig}

\let\tablenum\relax
\usepackage[exponent-mode=scientific]{siunitx}
\usepackage[version=4]{mhchem}
\newcommand{\mc}{\si{\micro\meter}}
\newcommand{\threeum}{\qty{3.1}{\mc}}
\newcommand{\fourmic}{\qty{4.57}{\mc}}

\accepted{In PSJ}

\graphicspath{{./}{figures/}}
\begin{document}

\title{Callisto from JWST: \ce{CO2}-rich terrain on the leading hemisphere and global patterns of \ce{H2O} ice}

\email{mcamarca@caltech.edu}
\author[0000-0003-3887-4080]{Maria Camarca}
\affiliation{Division of Geological and Planetary Sciences, California Institute of Technology, CA, USA}

\author[0000-0002-9068-3428]{Katherine de Kleer}
\affiliation{Division of Geological and Planetary Sciences, California Institute of Technology, CA, USA}

\author[0000-0002-6886-6009]{Richard J. Cartwright}
\affiliation{Johns Hopkins University Applied Physics Laboratory, MD, USA}

\author[0000-0002-2662-5776]{Geronimo L. Villanueva}
\affiliation{NASA Goddard Space Flight Center, MD, USA}

\author[0000-0002-6117-0164]{Bryan J. Holler}
\affiliation{Space Telescope Science Institute, MD, USA}

\author[0000-0001-5683-0095]{Zachariah Milby}
\affiliation{Division of Geological and Planetary Sciences, California Institute of Technology, CA, USA}

\author[0000-0002-3225-9426]{Kevin P. Hand}
\affiliation{Jet Propulsion Laboratory, California Institute of Technology, CA, USA}

\author[0000-0003-0554-4691]{Lorenz Roth}
\affiliation{Space and Plasma Physics, KTH Royal Institute of Technology, Stockholm, Sweden}



\begin{abstract}
We present maps of \ce{H2O}, \ce{CO2}, and a \fourmic{} spectral feature across Callisto's surface observed using \edit1{the James Webb Space Telescope (JWST)}. \ce{H2O} ice was mapped by measuring band parameters of the \threeum{} Fresnel peak across the leading and trailing hemispheres under a simplified assumption of crystalline ice. We update the \ce{CO2} solid-phase, \ce{CO2} gas, and \fourmic{} feature band depth maps originally presented in \cite{cartwright-2024-RevealingCallistos} with a new JWST observation of Callisto centered on Valhalla, the largest multi-ring impact basin in the solar system. Our \ce{H2O} ice map shows that the Fresnel peak on the trailing hemisphere exhibits a bullseye pattern that is weaker at low latitudes and on the leading hemisphere its strength is associated with impacts. This dichotomy is possibly related to the Jovian magnetospheric plasma impinging on the trailing hemisphere. Our solid-phase \ce{CO2} map reveals an enhancement in the vicinity of the Lofn/Heimdall impact craters, a region that may be the largest reservoir of non-radiolytic \ce{CO2} on Callisto's surface. The gas-phase \ce{CO2} exhibits a patchy spatial distribution and does not clearly correlate with solid \ce{CO2}. 

\end{abstract}



\section{Introduction}

The Galilean moons of Jupiter---Io, Europa, Ganymede, and Callisto---host an incredible treasury and range of geologic activity. Evidence for endogenic geologic processes including tectonic deformation, volcanism, and surface-interior interactions is readily found on Io \citep{morabito-1979-DiscoveryCurrently,dekleer-2024-Isotopicevidence,davies-2006-heartbeatvolcano}, Europa \citep{mccord-1999-Hydratedsalt,schmidt-2011-Activeformation}, and Ganymede \citep{bagenal-2004-JupiterPlanet}, but is largely absent from Callisto \citep{moore-2004-Callisto}. Although Callisto is excluded from the Laplace resonance that leads to the tidal heating on the inner three moons, \emph{Galileo} magnetometer results indicate the signature of an induced magnetic field, possibly originating from a saline subsurface ocean \citep{zimmer-2000-SubsurfaceOceans,cochrane-2025-StrongerEvidence,strack-2024-SpatiotemporalStructure,hartkorn-2017-Inductionsignals}. At the global scale, Callisto's geology is dominated by large impacts, including exotic multi-ring impact basins of great diameters unseen on any other solar system object \citep{moore-2004-Callisto}. Callisto's surface volatile inventory includes \ce{H2O}, \ce{CO2}, and possibly S-bearing compounds \citep{hibbitts-2000-DistributionsCO2,mccord-1998-Nonwatericeconstituents,cartwright-2020-EvidenceSulfurbearing,carlson-1999-TenuousCarbon}. These volatiles are accompanied by the largest share of dark material in the Galilean system, as evidenced by Callisto bearing the lowest albedo ($\sim$0.2) of the four moons (e.g., see Fig. 4 of \citealt{king-2025-SpatiallyResolved}). This dark material is hypothesized to be a thick lag deposit formed by the removal of high-albedo volatiles via continual sublimation \edit1{\citep{moore-1999-MassMovement,spencer-1987-Thermalsegregation,spencer-1984-Mobilitywater}}, akin to a similar process observed on comets. Substantial amounts of dust may have accumulated on Callisto's surface over the age of the solar system, perhaps with some of it originating from the Jovian irregular satellites \citep{bottke-2013-Blackrain}. As such, by virtue of its geologic quiescence, Callisto serves as a template for understanding how \edit1{icy moons evolve} under the near exclusive control of exogenic deposition and irradiation processes. 

Exposed \ce{H2O} ice on Callisto and its sibling moons is central to feeding surface chemistry and atmospheric processes. Because the Galilean satellites are embedded in the intense Jovian magnetosphere, \ce{H2O} ice on their surfaces is altered by charged particles. \edit1{Such particles include electrons, protons, and oxygen and sulfur ions of varying charge states (e.g., S\textsuperscript{++}), spanning a wide range of energies from \edit1{$\sim$10s eV} to 100s of MeV (\citealt{cassidy-2010-RadiolysisPhotolysis,bagenal-2015-Plasmaconditions} and references therein)}. For instance, \ce{H2O} ice radiolysis is a fundamental step in the formation of molecular oxygen \citep{johnson-1997-Photolysisradiolysis}.  Molecular oxygen exists in the icy Galilean satellite atmospheres as inferred from UV and optical atomic O emissions \citep{dekleer-2023-OpticalAurorae,cunningham-2015-DetectionCallistos,hall-1995-Detectionoxygen,hall-1998-FarUltravioletOxygen,roth-2016-Europasfar}, as well as on their surfaces, likely trapped in H$_2$O ice \citep{spencer-2002-CondensedO2,trumbo-2021-GeographicDistribution,oza-2024-Commonorigin,johnson-1997-O2O3}. \edit1{Additionally, \ce{H2O} ice might function as a trapping mechanism for some portion of solid-phase \ce{CO2} on icy satellites (e.g., \citealt{cruikshank-2010-Carbondioxide,brown-2025-JWSTStudy})}. In the near-infrared, the state of \ce{H2O} ice may be probed through analyses of spectral features in the 1 to 3 \mc{} range. \edit1{Examples of observations or inferences of \ce{H2O} ice on Callisto go back many decades (e.g., \citealt{lebofsky-1977-Identificationwater,pollack-1978-Nearinfraredspectra,clark-1980-GalileanSatellites,pilcher-1972-GalileanSatellites,morrison-1973-Thermalproperties,fink-1973-InfraredSpectra}}), and one of the primary sources of information about Callisto's \ce{H2O} ice was the \edit1{Near-Infrared Mapping Spectrometer (NIMS)} instrument onboard the \emph{Galileo} spacecraft. The NIMS instrument provided spectral measurements in the 0.7--5.2 \mc{} range with a spectral resolving power of R $\sim$40--200 \citep{carlson-1992-NearInfraredMapping}. Generally, the NIMS Callisto observations showed that the presence of exposed \ce{H2O} ice generally tracked with the presence of craters, however the spectra were often noisy due to internal instrument performance and Callisto's low levels of exposed surface ice ($\sim$10 \% areal coverage, \citealt{hansen-2004-Amorphouscrystallinea}).  

\edit1{Beyond \ce{H2O} ice, Callisto's volatile library also retains \ce{CO2}, which is notable given that \edit1{this compound is not stable at the orbit of Jupiter \citep{watson-1962-stabilityvolatiles, ahrens-2022-GeoscientificReview}}}. Indeed, many surfaces in the Jovian system bear solid \ce{CO2}, including the icy Galilean moons \citep{cartwright-2024-RevealingCallistos,mccord-1998-Nonwatericeconstituents,hibbitts-2002-CO2richimpact,villanueva-2023-EndogenousCO,trumbo-2023-distributionCO2} as well as the irregular satellite \citep{sharkey-2025-JWSTReveals} and Trojan populations \citep{wong-2024-JWSTNearinfrared}. Although pure \ce{CO2} ice as studied in the lab has an asymmetric stretch ($\nu_3$) absorption feature at 4.27 \mc{} \citep{sandford-1990-PhysicalInfrared}, measured icy moon \ce{CO2} often presents a shorter wavelength feature closer to 4.25 \mc{} indicative of trapped \ce{CO2}, with proposed host materials including both icy (e.g., \ce{H2O}) and non-icy substrates \citep{brown-2025-JWSTStudy,ahrens-2022-GeoscientificReview}. As such, the origin of Jovian satellite \ce{CO2} may be revealed by placing the spatial distribution of the \ce{CO2} band shape and center in context with geologic features. For example, the enhanced abundance of Europa's surficial \ce{CO2} near chaos terrains, combined with an analysis of the $\nu_{3}$ band shape, strongly suggests a link to recent, internal sources \citep{trumbo-2023-distributionCO2,villanueva-2023-EndogenousCO}. On Callisto and Ganymede, their surface \ce{CO2} further acts to feed their \ce{CO2} atmospheres \citep{carlson-1999-TenuousCarbon,cartwright-2024-RevealingCallistos,bockelee-morvan-2024-patchyCO2}. However, the balance of mechanisms that promote the release of \ce{CO2} from the surface into the atmosphere remains uncertain. As demonstrated by \cite{cartwright-2024-RevealingCallistos}, the peak column densities of \ce{CO2} gas on Callisto do not align with the peak \ce{CO2} surface abundance as inferred from band depth maps, nor do they align with the equatorial afternoon where the surface temperature is warmest. More complete coverage of the \ce{CO2} gas would help in determining how universal this patchiness is across Callisto's surface.

At present, studies of \ce{H2O} and the chemistry of volatiles in the Galilean system are being greatly advanced by the James Webb Space Telescope (JWST). As an infrared telescope, JWST is sensitive to the ro-vibrational absorption and emission features of solid- and gas-phase volatiles that dominate the surface compositions of these moons. With the incredible spatial resolution of JWST, the location of these volatiles can be tethered to geologic maps of the surface made from spacecraft data, such that their origins (either endogenic or exogenic) can be evaluated. For example, SO gas has been observed directly above an active volcano on Io \citep{depater-2023-EnergeticEruption}, and observations of Ganymede show in great detail how the moon's intrinsic magnetic field promotes the collection of select volatiles near the poles \citep{trumbo-2023-Hydrogenperoxide,bockelee-morvan-2024-patchyCO2}. 

The first published analysis of Callisto using JWST's Near Infrared Spectrograph instrument (NIRSpec) focused on one observation each of the leading and trailing hemispheres \citep{cartwright-2024-RevealingCallistos}. This analysis offered a comprehensive look at Callisto's abundant surface \ce{CO2}, and included the first detection of \ce{CO2} gas on Callisto since the \edit1{\emph{Galileo} mission} in the 1990s \citep{carlson-1999-TenuousCarbon}. Here, we analyze a third JWST observation of Callisto centered on its largest impact basin, Valhalla. Using these data, we provide new full-coverage maps of \ce{H2O} ice  using the \threeum{} Fresnel reflectance peak, as well as updated analysis of \ce{CO2} and a possible \fourmic{} organics feature explored in \cite{cartwright-2024-RevealingCallistos}. Within the existing JWST data set for Callisto, the 3.1 \mc{} Fresnel peak is the sole \ce{H2O} ice feature that is fully sampled across wavelength space. Although this feature is sensitive to various attributes of water ice (e.g., grain size, temperature), determining how its major properties \edit1{such as band depth and band center vary} across the disk offers a simplified overview of crystalline \ce{H2O} \edit1{ice} processes on Callisto.

The organization of this paper is as follows: Section~\ref{jwst-methods} describes the reduction of the JWST NIRSpec data and parameter extraction for the spectral features. Section~\ref{jwst-results} describes the results of mapping the Fresnel Peak across Callisto, as well as the maps of \ce{CO2} and the \fourmic{} feature. Section~\ref{jwst-discussion} offers our interpretation of these features, including how localized enhancements do/do not correlate with local geology. We also place these new Callisto results in context with those obtained for the other icy Galilean moons. Section~\ref{jwst-conclusion} summarizes the conclusions of this work.

\section{Methods}\label{jwst-methods}

\subsection{Observations and Data Reduction}
\begin{figure}
\centering
\includegraphics[width=0.5\textwidth]{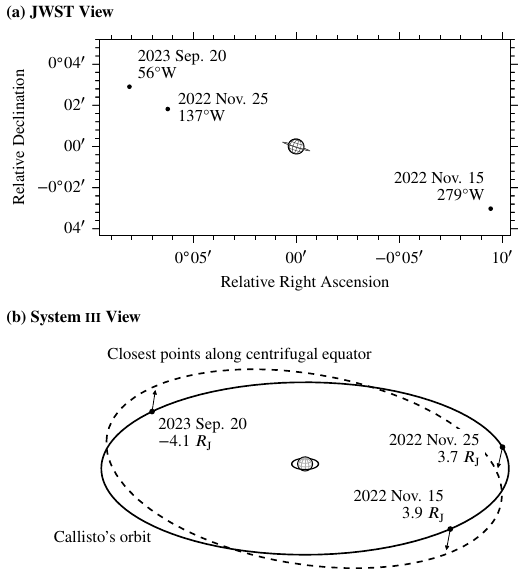}
\caption{(a) The projected separation of Callisto from Jupiter at the time of each JWST observation from the perspective of the telescope. The sub-observer longitude of Callisto is shown for each date. (b) The position of Callisto relative to the co-rotating Jovian plasma sheet (dashed line) at the time of each observation, in the reference frame of the plasma sheet. The arrows are scaled for both direction and relative distance. The labeled dates are accompanied by the distance of Callisto from the middle of the sheet \edit1{in Jupiter radii, $R_{J}$ = 71492 km \citep{archinal-2018-ReportIAU} }. With the exception of the physical size of Callisto, the drawing is to scale. In this visualization, the direction of sheet motion is counter clockwise.}
\label{jwst-fig:geometry}
\end{figure}

Observations of Callisto were obtained using the James Webb Space Telescope (JWST), located at the $L_2$ Lagrange point of the Sun-Earth system \citep{gardner-2023-JamesWebb}. With its 6.5 m primary mirror, excellent sensitivity, and high spatial resolution, JWST enables detailed mapping of spectral features on solar system targets. We used the NIRSpec instrument \citep{jakobsen-2022-NearInfraredSpectrograph} as part of General Observer program 2060 (P.I. Cartwright), which observed Callisto on 2022 November 15 and 25 (previously analyzed by \citealt{cartwright-2024-RevealingCallistos}) and 2023 September 20 (newly presented in this work). The respective sub-observer longitudes of these observations were \ang[exponent-mode=input]{279}W, \ang[exponent-mode=input]{137}W, and \ang[exponent-mode=input]{56}W; these viewing geometries sampled Callisto's trailing hemisphere, the Asgard impact basin, and the Valhalla impact basin, respectively. The position of Callisto relative to Jupiter, and relative to Jupiter's centrifugal equator, are shown in Fig.~\ref{jwst-fig:geometry}.  Data were collected using NIRSpec’s integral field unit (IFU) that has a \SI{3}{\arcsecond} $\times$ \SI{3}{\arcsecond} field of view, and a pixel size of 0.1$\arcsec$. The G395H grating was used in all three observations, spanning 2.85--5.35 \mc{} with an average resolving power R $\sim$ 2700.  Each observation was acquired via a set of four 32 s exposures which each sampled different parts of the detector (four dithers), using the NRSRAPID readout mode. 

The data were downloaded from the Mikulski Archive for Space Telescopes (MAST) at the Space Telescope Science Institute, and they can be accessed via \doi{10.17909/e61x-e585}. The JWST pipeline versions used for each observation were: 1.12.5 (Sep 20), 1.7.2 (Nov 15), and 1.8.2 (Nov 25). The Calibration References Data System files used for data taken on Nov 15, Nov 25, and Sep 20 respectively are: \verb|jwst_1015.pmap|, \verb|jwst_1017.pmap|, and \verb|jwst_1169.pmap|. Additional custom codes were developed to combine dithered frames and remove bad pixels. The four dithers were georeferenced to Callisto’s disk and then median combined, which assisted in removing abnormal pixels. At each spatial pixel (spaxel), we distinguished between reflectance and thermal emission components by applying a two-component model: a realistic solar model for reflected light and a Planck function for thermal radiation. We generated the solar model using the Planetary Spectrum Generator (PSG; \citealt{villanueva-2018-PlanetarySpectrum,villanueva-2022-FundamentalsPlanetary}), which adjusts for Doppler shifts, uses the high-fidelity ACE solar spectrum (\citealt{hase-2010-ACEFTSatlas}) to account for solar Fraunhofer lines, and employs the \citet{kurucz-2005-Newatlases} model to reproduce the continuum intensity. After subtracting the thermal component, we calculated the reflectance spectrum for each spaxel by dividing the calibrated flux values by a solar model. This model was scaled according to the spaxel's projected spatial dimensions and corrected for the Sun-Callisto and JWST-Callisto distances at the observation times. This methodology parallels that used in previous analyses of NIRSpec IFU data for Europa \citep{villanueva-2023-EndogenousCO}, Callisto \citep{cartwright-2024-RevealingCallistos}, and Enceladus \citep{villanueva-2023-JWSTmolecular}.

\subsection{Band Parameter Analysis}
\emph{Retrieval of solid-state band parameters:}
After the NIRSpec cubes were thermally-corrected and dither-averaged, we mapped the spatial distribution of spectral features of interest. First, we registered the center point of Callisto's disk and determined the range of pixels that fall within the projected radius of Callisto at the time of each observation. After spatial registration, we proceeded to map the 4.25 \mc{} \ce{CO2} feature, the \threeum{} \ce{H2O} ice Fresnel peak, and a \fourmic{} absorption feature. For these fits, we used the Python package \texttt{lmfit} to retrieve band parameters and statistics \citep{newville-2025-LMFITNonLinear}. \edit1{For the 4.25 \mc{} \ce{CO2} feature, we simultaneously fit a two-Gaussian absorption profile and a linear continuum, while for the \fourmic{} band we used a single Gaussian plus a linear continuum.} The maximum absorption depth for each feature was divided by the local continuum value to retrieve the fractional band depth. For the \threeum{} Fresnel reflectance peak, we fit a three-component model that consisted of a 1) linear continuum from 3.00 to 3.30 \mc{}, 2) a Gaussian profile centered near 3.1 \mc{}, and 3) a second Gaussian profile centered near 3.2 \mc{}. We initially tested the Fresnel peak fits with a single Gaussian at \threeum, however the need for a smaller, secondary component was motivated by structure apparent in the residuals. The multi-peaked structure of the Fresnel peak is well-documented in icy satellites (including Callisto, \citealt{hansen-2004-Amorphouscrystallinea}). \edit1{We summed the areas of both peaks above the continuum to determine the total band area in units of \mc{}. We note this approach differs from the Fresnel peak band measurement techniques used for Ganymede NIRSpec data by \citep{bockelee-morvan-2024-Compositionthermal,trumbo-2023-Hydrogenperoxide} and for Europa \citep{cartwright-2025-europa-fresnel}, for which the reported band areas are normalized by the underlying continuum. However, \edit1{as a reflection off ice crystal facets}, and not absorption or emission feature, the Fresnel peak is not a function of the underlying continuum, which can be influenced by both water ice and non-ice minerals. At least for Callisto, we verified that its Fresnel peak distribution is similar between the two methods. For the Fresnel peaks extracted from individual pixels, we find typical errors of $\pm$\SI[scientific-notation=fixed, fixed-exponent=-4,round-mode=figures, round-precision=1]{7.318e-06}{\mc} for the total band area, and $\pm$0.0007 $\mc$ for the band center of the main peak. Likewise, for the 4.25 \mc{} \ce{CO2} feature, we calculate typical uncertainties of $\pm$ 0.4$\%$ and  $\pm$0.0006\mc{} for the band depth and center, respectively. Regarding the \fourmic{} feature, the individual pixel uncertainties are $\pm$ 0.15$\%$ for the band depth and $\pm$0.001 $\mc$ for the band center. }

\emph{Retrieval of gaseous \ce{CO2}:} 
Although Callisto's spectral profile around 4.25 \mc{} is dominated by absorption from the solid-phase \ce{CO2}, weak  gaseous features can be extracted from this same wavelength range via a careful treatment of the continuum. This procedure for JWST NIRSpec observations was first demonstrated at Callisto \citep{cartwright-2024-RevealingCallistos}, and has since been applied to Ganymede \citep{bockelee-morvan-2024-patchyCO2}. Here, we reiterate the primary details of this extraction. Excited molecules of \ce{CO2} gas produce a fluorescent emission band in the 4.2 to 4.3 \mc{} spectral range. \edit1{This band is composed of two groups of emission lines called the P and R branches that correspond to different rotational quantum number changed associated with the transitions \citep{bosman-2017-CO2infrared} }. These narrow emission lines that make up these branches are resolvable at the resolution of NIRSpec's G395H grating (R $\sim$ 3000 at 4.25 \mc). These lines are separated from the solid-phase \ce{CO2} absorption by generating a smoothed Callisto spectrum between 4.2 and 4.3 \mc{} (R = 1000) that only retains the shape of the solid-phase profile. This smoothed continuum profile was then subtracted from the native resolution NIRSpec data (R $\sim$ 3000), to isolate the residual CO$_2$ gas emission features. We performed this technique on all spaxels covering Callisto’s disk and a $\sim$\ang[exponent-mode=input]{;;0.3} wide annulus of spaxels beyond its disk to search for \ce{CO2} gas over a range of altitudes above its surface (up to $\sim$1000 km). Next, the residuals containing the gaseous emission lines were cross-correlated to synthetic spectra of varying \ce{CO2} concentrations generated by PSG. The output of this approach is an estimate of the average line-of-sight \ce{CO2} gas column density as seen by the observer. This line-of-sight was not corrected for the observing geometry (i.e., incidence and emission angles). For our PSG model, we assume that the excitation process is photon-dominated, rather than electron-dominated. An assessment of \ce{CO2} electron excitation, employing an electron population comparable to that used for analyzing auroral emissions \citep{cunningham-2015-DetectionCallistos,dekleer-2023-OpticalAurorae}, indicates minimal contributions (\textless0.1 \%), reinforcing solar-pumped fluorescence as the primary excitation mechanism at Callisto. The observation rms and $\chi^{2}$ statistics between the finalized continuum model and the data informed the 1$\sigma$ uncertainties for the resulting \ce{CO2} column density estimates.

\section{Results}\label{jwst-results} 

\begin{figure}
\centering
\includegraphics[width=1.0\textwidth]{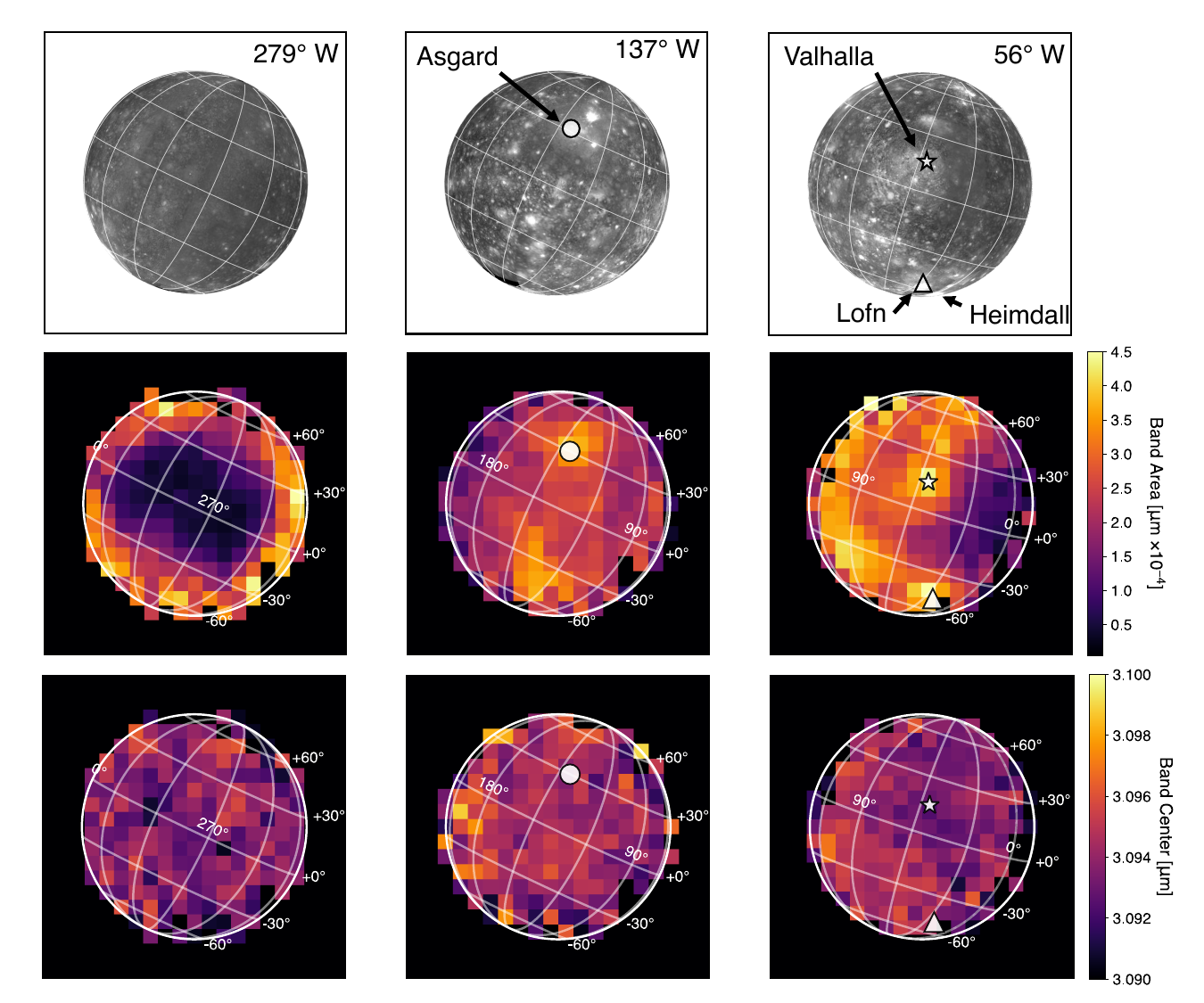}
\caption{The band areas and centers of Callisto's \threeum{} water ice Fresnel peak for three viewing geometries. The top row shows projected USGS albedo maps of Callisto stretched to allow for easier identification of the major geologic terrains. The middle row shows the band areas, and the bottom row shows the band centers. \edit1{The approximate centers of the following geologic units are marked with symbols: Valhalla, $\star$ ; Asgard, $\circ$; Lofn/Heimdall, $\triangle$.} The sub-observer longitude of each observation is indicated in the panels. The projected angular size of Callisto varies slightly between each observation, so the spaxel footprint is slightly different in each panel. \edit1{Typical errors are $\pm$\SI[scientific-notation=fixed, fixed-exponent=-4,round-mode=figures, round-precision=1]{7.318e-06}{\mc} for the band area and $\pm$0.0007 $\mc$ for the band center.} }
\label{jwst-fig:fresnel-maps}
\end{figure}

\subsection{\edit1{Variation} in the Fresnel Peak across Callisto}

\begin{figure}
\centering
\includegraphics[width=0.3\textwidth]{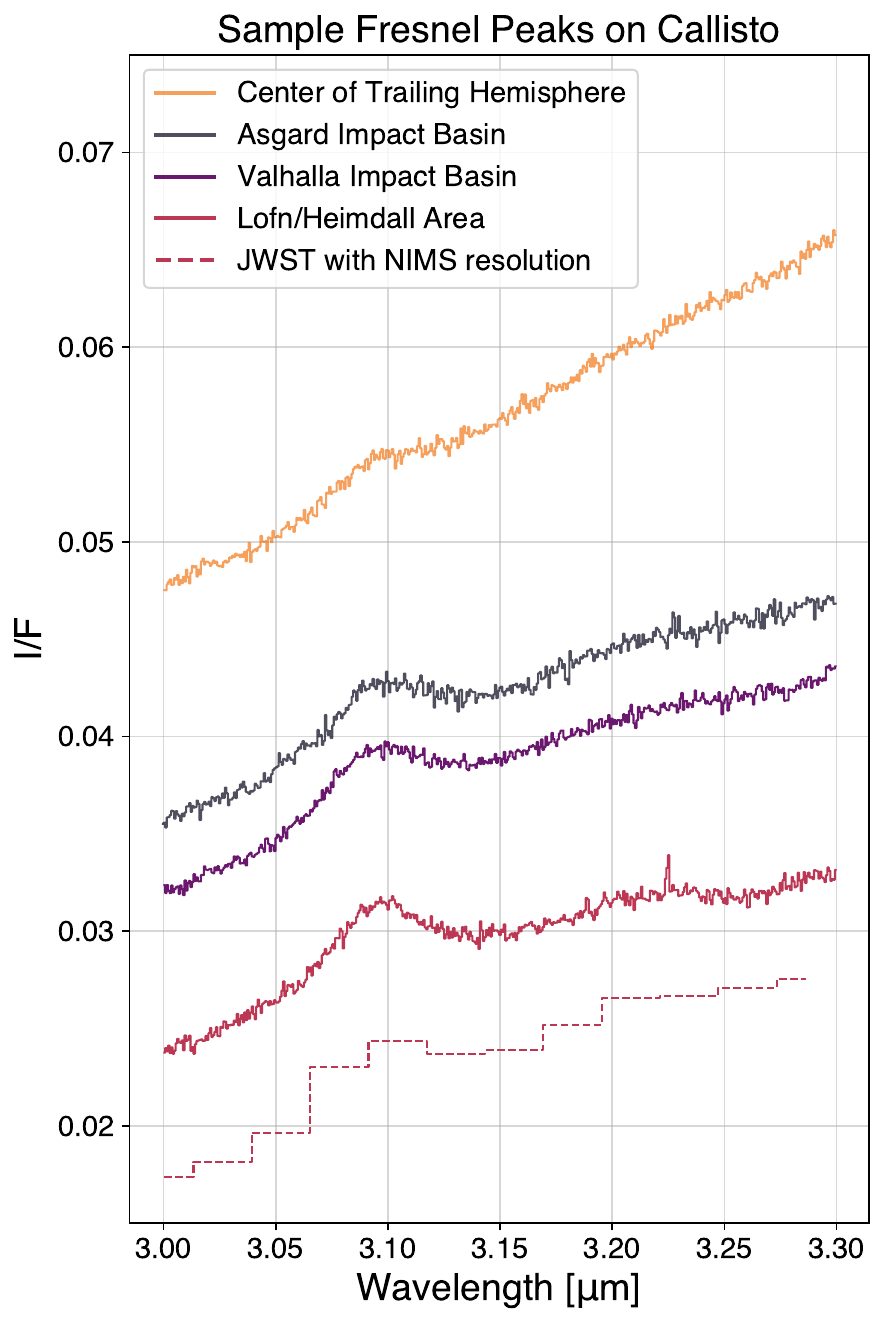}
\caption{A sample of Fresnel Peak spectra taken from the three JWST NIRSpec observations of Callisto. Each spectrum represents data from one representative spaxel for each of the locations listed. The spectra are shown prior to continuum subtraction. The Fresnel peak is smaller and more muted across most of Callisto's trailing hemisphere. On the leading hemisphere, spaxels that overlap with Callisto's major multi-ring basins, Asgard and Valhalla, show more defined structure. A spaxel from the Lofn/Heimdall region shows a particularly large Fresnel peak for Callisto. A comparison of a Fresnel peak from a JWST spaxel with that same data sampled every $\sim$0.026 \mc{} to simulate the resolution of NIMS near 3 \mc{} \citep{hansen-2004-Amorphouscrystallinea}. This basic comparison only serves to showcase the excellent resolution of JWST compared to NIMS, not the sensitivity difference.}
\label{jwst-fig:fresnel-spectra}
\end{figure}

\edit1{The band area maps of Callisto's Fresnel peak as presented in Fig.~\ref{jwst-fig:fresnel-maps} demonstrate that crystalline \ce{H2O} ice is not homogeneous across the moon. Instead, this \ce{H2O} ice feature exhibits leading vs. trailing hemisphere differences, as well as localized enhancements. To complement the Fresnel peak maps, Fig.~\ref{jwst-fig:fresnel-spectra} shows sample spectra from the three observations to highlight the improved spectral resolution of JWST NIRSpec compared to \emph{Galileo} NIMS.}

\emph{Leading vs. Trailing Hemisphere}: The Fresnel peak is clearly different on the trailing hemisphere in both strength and spatial distribution compared to the leading hemisphere (Fig.~\ref{jwst-fig:fresnel-maps}). We find band areas are larger on Callisto's leading hemisphere. \edit1{The average trailing hemisphere band area is $\sim$\SI[scientific-notation=true]{0.00019}{\mc}, while typical values for the Asgard and Valhalla observations are $\sim$1.2 and $\sim$1.3 times this value, respectively (Fig.~\ref{jwst-fig:fresnel-maps}).} Additionally, the trailing hemisphere exhibits a distinct bullseye pattern with weaker band strengths at low latitudes, and increasing band strengths at high latitudes and toward the anti-Jovian and sub-Jovian limbs. The Asgard and Valhalla observations do not exhibit this bullseye pattern. Instead, the largest band areas on the two leading hemisphere observations \edit1{are generally associated with large impacts}. 

\emph{Localized Enhancements on Asgard and Valhalla}:
For the Asgard-centered observation, there are two main localized regions of elevated band areas. One is centered near the Asgard impact basin at \ang[exponent-mode=input]{32}N, \ang[exponent-mode=input]{140}W. A representative value for this area is $\sim$\SI[scientific-notation=true]{0.00037}{\mc{}}, which is $\sim$1.4 times greater than neighboring pixels at similar latitudes. Another smaller (D$\sim$610 km) multi-ring basin, Utgard, is just north of Asgard's center at \ang[exponent-mode=input]{45}N, \ang[exponent-mode=input]{134}W and may also be contributing to this same group of elevated band strength. The second area of elevated Fresnel peak strength in this observing geometry is in the southern hemisphere, \edit1{around the sub-observer longitude}. Beyond these two regional enhancements, the Fresnel band area measurements across the Asgard-centered observation are fairly uniform. 

Similarly, we find additional enhancements of the Fresnel peak on the Valhalla-centered observation. The dominant feature in the center of Callisto's disk for this observation is the increased band area near the center of Valhalla, located at \ang[exponent-mode=input]{14}N, \ang[exponent-mode=input]{56}W. The band areas of pixels near the center are $\sim$\SI[scientific-notation=true]{0.0004}{\mc{}}. In the south, we find larger swaths of higher band areas, including one region that appears co-located with Lofn near \ang[exponent-mode=input]{-56}N, \ang[exponent-mode=input]{22}W. \edit1{The corresponding reflectance spectra for these regions are shown in Fig.~\ref{jwst-fig:fresnel-spectra}, which illustrates the more pronounced Fresnel peak structure associated with the impact basins, especially compared to the trailing hemisphere}.  Beyond these broad enhancements of the Fresnel peak area, we also identify a region east of Valhalla between ($\sim$\ang[exponent-mode=input]{0}--\ang[exponent-mode=input]{30}N, \ang[exponent-mode=input]{0}--\ang[exponent-mode=input]{30}W) that bears generally weaker band areas than the pixels to the west of Valhalla. The Fresnel peak areas in this region are $\sim$0.5$\times$ lower than the Valhalla-observation full-disk average.

\subsection{\texorpdfstring{The 4.25 \textmu{m} \ce{CO2} Feature across Callisto}{The 4.25 \textmu m CO2 Feature across Callisto}}

\begin{figure}
\centering
\includegraphics[width=1.0\textwidth]{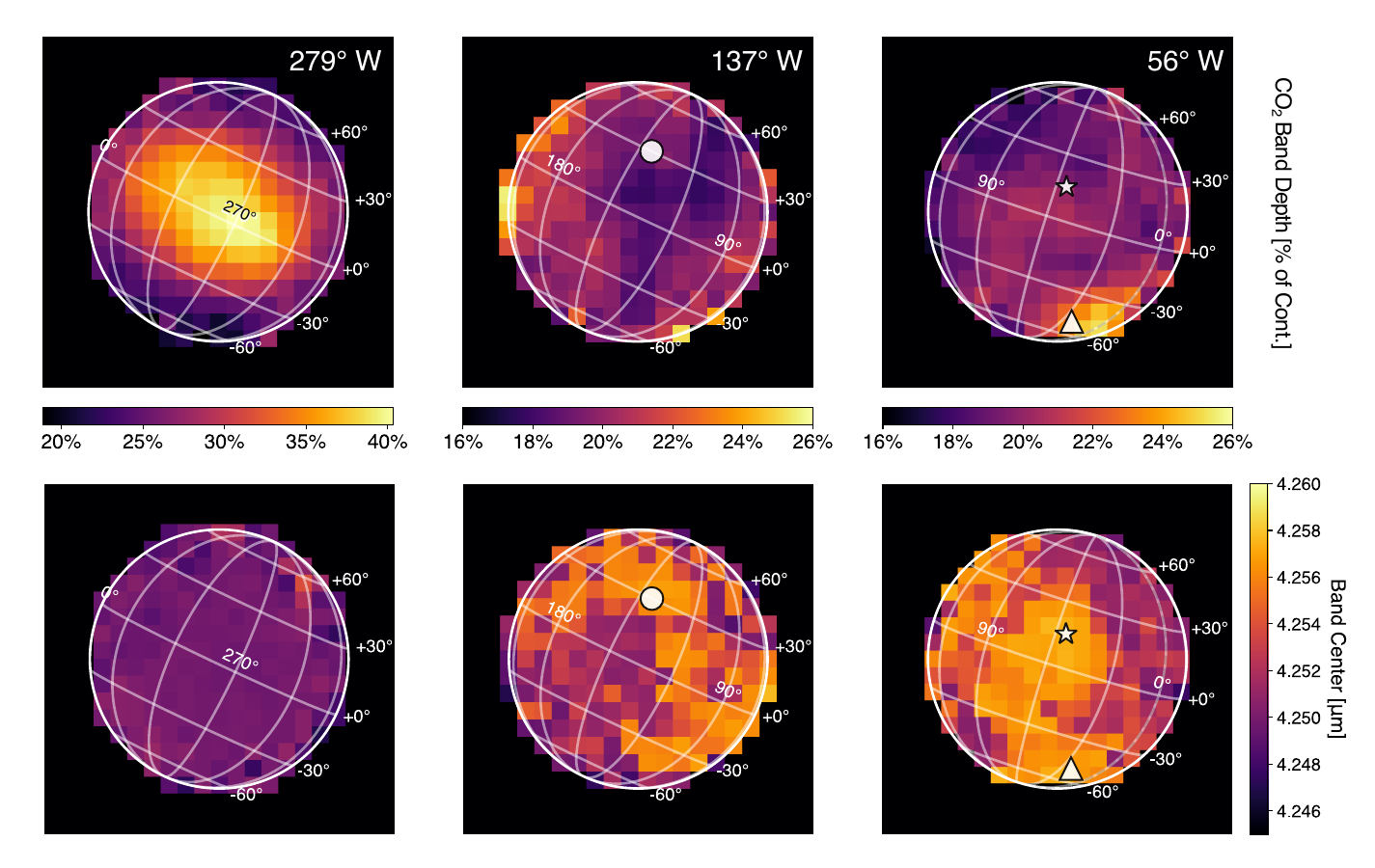}
\caption{The band depths and centers of Callisto's \ce{CO2} as measured at 4.25 \mc{} for three viewing geometries. The trailing hemisphere and Asgard-centered data were previously published by \citet{cartwright-2024-RevealingCallistos}. \edit1{The following geologic units are marked with symbols: Valhalla, $\star$ ; Asgard, $\circ$; Lofn/Heimdall, $\triangle$.} The sub-observer longitude of the observation is indicated in the panels. The projected angular size of Callisto varies slightly between each observation, so the spaxel scale is slightly different in each panel. The typical uncertainty in the band depth is $\pm$ 0.4$\%$, while the band center is $\pm$0.0006\mc{} }
\label{jwst-fig:co2-solid}
\end{figure}

\edit1{In Fig.~\ref{jwst-fig:co2-solid}, we present an updated map of the 4.25 \textmu{m} solid \ce{CO2} solid feature across Callisto that includes the Valhalla impact basin.} The disk-averaged \ce{CO2} band depth is $\sim$19 \%, while the enhanced pixels near \ang[exponent-mode=input]{0}W and \ang[exponent-mode=input]{-45}N are closer to $\sim$25 \%. These \ce{CO2}-rich pixels are geographically near Callisto's most prominent southern hemisphere impact features, mainly Lofn and Heimdall (\ang[exponent-mode=input]{-63}N, \ang[exponent-mode=input]{357}W). In contrast to the Fresnel peak data, the \edit1{solid-phase \ce{CO2} distribution around Valhalla appears more diffuse}.

\begin{figure}
\centering
\includegraphics[width=1.0\textwidth]{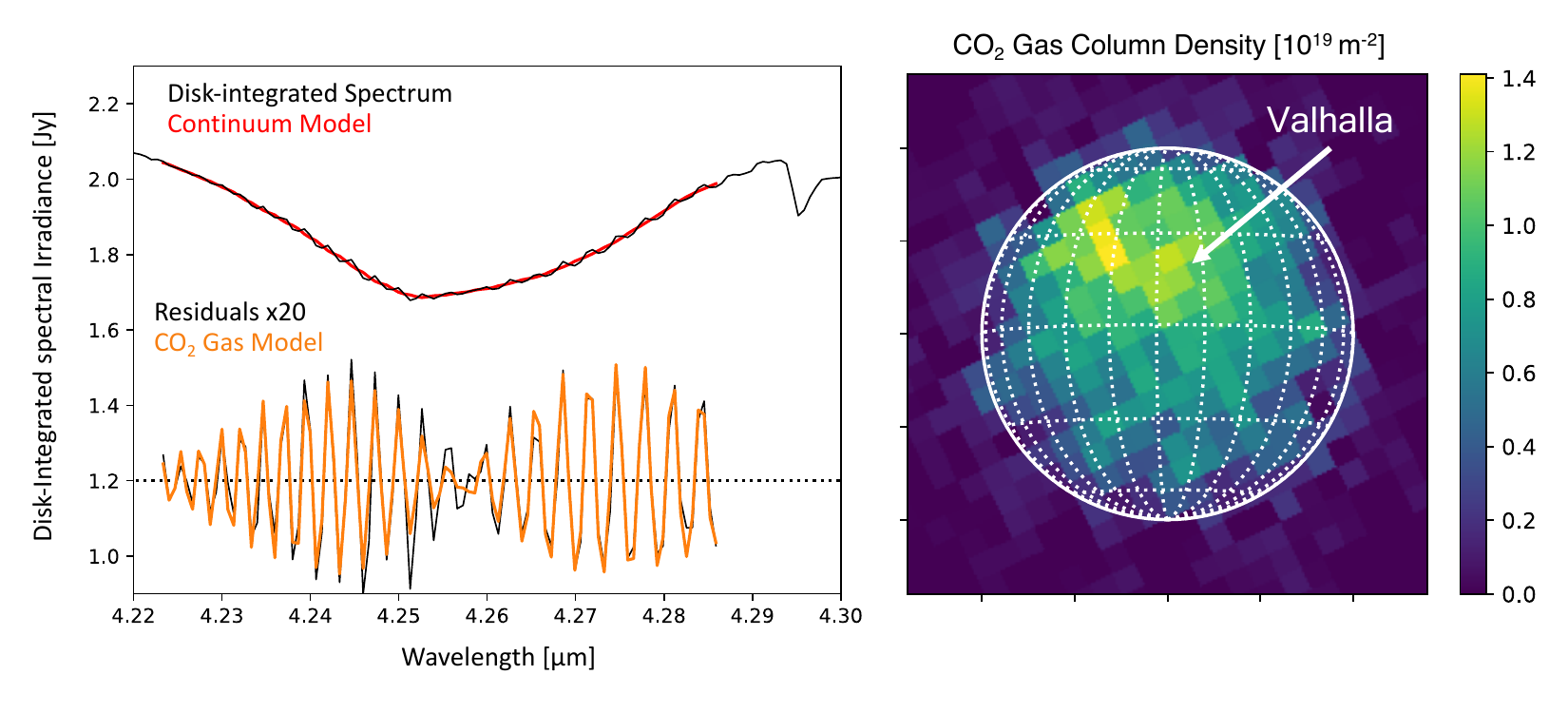}
\caption{A detection of \ce{CO2} gas on the Valhalla-centered observation using JWST. \emph{Left panel:} A slice of the disk-integrated NIRSpec spectrum containing the 4.25 \mc{} absorption feature (black). This absorption feature is dominated by solid-phase \ce{CO2} to which a continuum was fit (red). The subtraction of the continuum from the data reveals the much weaker gas phase signal (black, bottom). The fit of the \ce{CO2} gas-phase model is shown in orange. \emph{Right panel:} The column density of \ce{CO2} gas across Callisto's disk.}
\label{jwst-fig:co2-gas}
\end{figure}

\edit1{In addition to detecting solid-phase \ce{CO2} \edit1{across the Valhalla-centered hemisphere}, Fig.~\ref{jwst-fig:co2-gas} shows the detection of gas phase \ce{CO2}.} The characteristic sawtooth pattern of the $\nu_3$ fundamental vibration of gaseous \ce{CO2} is seen in the residuals (left panel of Fig.~\ref{jwst-fig:co2-gas}). As seen in the right panel of Fig.~\ref{jwst-fig:co2-gas}, the NIRSpec spaxels with the highest column densities (\textgreater \SI{1.1e19}{m^{-2}}) \edit1{ are in the northern hemisphere, within/near Valhalla}. \edit1{We report the mean column densities and corresponding standard deviations of \ce{CO2} for the trailing, Asgard-, and Valhalla-centered observations of \SI[scientific-notation=fixed, fixed-exponent=-19, round-precision=1, separate-uncertainty=true]{0.5 +- 0.2e-19}{}, \SI[scientific-notation=fixed, fixed-exponent=-19, round-precision=1, separate-uncertainty=true]{0.5 +- 0.2e-19}{}, and \SI[scientific-notation=fixed, fixed-exponent=-19, round-mode=figures, round-precision=1, separate-uncertainty=true]{0.6 +- 0.3e-19}{m^{-2}}, respectively.}

\subsection{The Distribution of the 4.57~\textmu{m} Feature}
\begin{figure}
\centering
\includegraphics[width=1.0\textwidth]{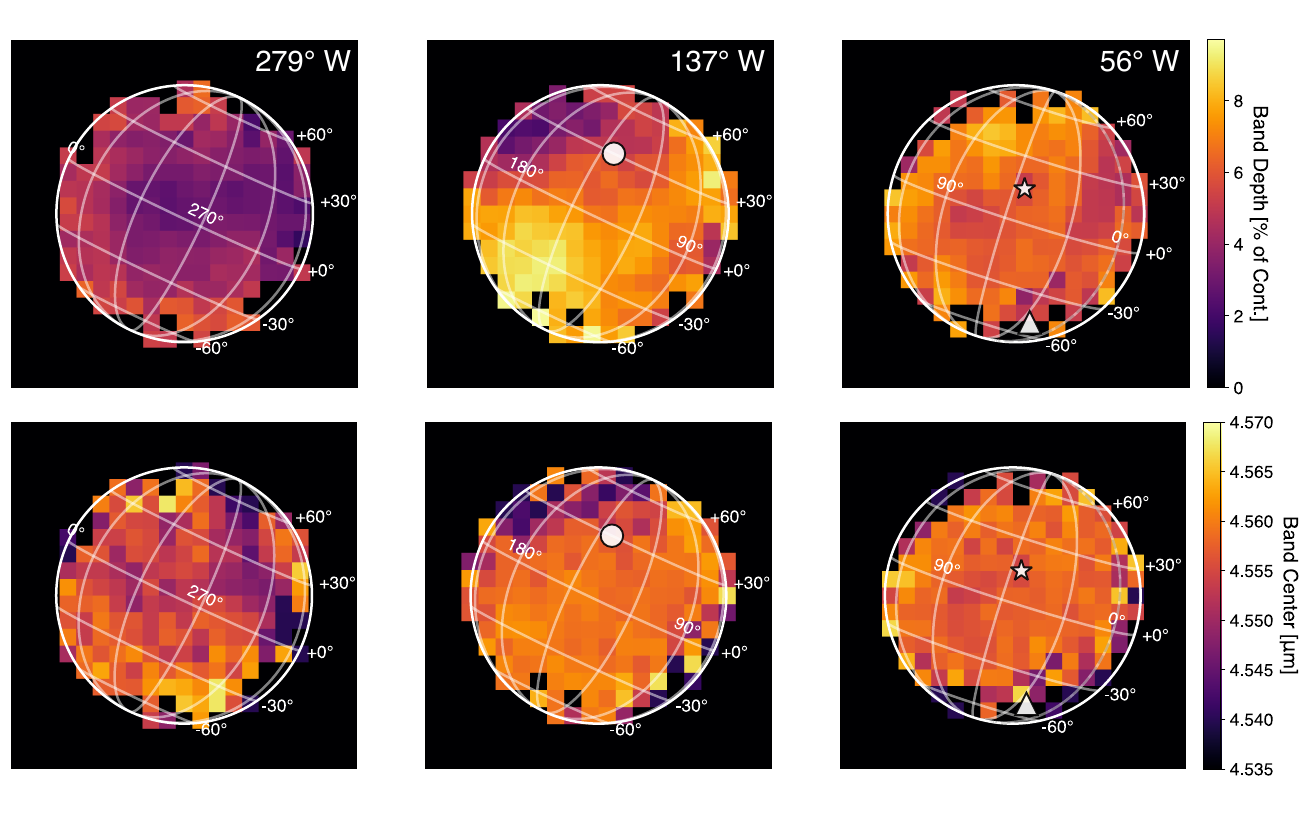}
\caption{The band depths and centers of Callisto's \fourmic{} feature for three viewing geometries. The trailing hemisphere and Asgard-centered data were previously published by \citet{cartwright-2024-RevealingCallistos}. \edit1{The centers of the following geologic units are marked with symbols: Valhalla, $\star$ ; Asgard, $\circ$; Lofn/Heimdall, $\triangle$.} The sub-observer longitude of the observation is indicated in the panels. The projected angular size of Callisto varies slightly between each observation, so the spaxel scale is slightly different in each panel.  \edit1{Typical uncertainties are $\pm$ 0.15$\%$ for the band depth and $\pm$ 0.001 $\mc$ for the band center. }}
\label{jwst-fig:456-results}
\end{figure}

\edit1{Beyond the maps of the \threeum{} Fresnel peak and the \SI{4.25}{\micro{m}} \ce{CO2} absorption feature on Callisto's Valhalla-centered observation, Fig.~\ref{jwst-fig:456-results} presents an update for the \fourmic{} spectral feature.} \edit1{The material responsible for the \fourmic{} is found all across the leading hemisphere, and the band depths remain the highest on the Asgard-centered observation. It appears that the \fourmic{} band depths may be slightly weaker near Valhalla relative to nearby terrain of similar latitude}. There is another possible region of weaker \fourmic{} band depths to the southwest of Valhalla's center, roughly between 0 to \ang[exponent-mode=input]{30}W and 0 to \ang[exponent-mode=input]{30}N.

\section{Discussion}\label{jwst-discussion}

\subsection{The Geologic Context of Callisto's 3.1~\textmu{m} Fresnel Peak}
In Section~\ref{jwst-results}, we presented a high-sensitivity, comprehensive map of Callisto's \threeum{} Fresnel feature. Because these Fresnel peaks are generated by the first reflection of light off an \ce{H2O} crystal \edit1{(e.g., \citealt{hansen-2004-Amorphouscrystallinea} and references therein)}, these maps can be interpreted as an approximation of the areal exposure of \ce{H2O} ice on Callisto's surface. However, it is important to emphasize the intensity, shape, and band center of the Fresnel peak are functions of multiple factors including temperature, grain size, and crystal structure (i.e., crystalline vs. amorphous, see  \citealt{stephan-2020-H2Oiceparticle,hansen-2004-Amorphouscrystallinea,mastrapa-2009-OPTICALCONSTANTS}). It is beyond the scope of the current work to include these factors in our analysis, as it would require detailed laboratory experiments and robust radiative transfer modeling. As such, we publish these Fresnel peak area maps as a ``first-pass'' at Callisto's \ce{H2O} ice distribution as observed by JWST, much like the approach taken for the JWST Ganymede Fresnel peak maps by \citet{bockelee-morvan-2024-Compositionthermal}. These ``crystalline'' results are complemented by recent work by \citet{cartwright-2025-europa-fresnel} on analyzing the state of \ce{H2O} ice across the surface of Europa using JWST.

The Fresnel maps presented here represent a step forward in quality compared to the prior \emph{Galileo} NIMS dataset. As illustrated in Fig.~\ref{jwst-fig:fresnel-spectra}, the better resolution of JWST allows for a higher fidelity extraction of features compared to the NIMS data. Because the \threeum{} signal in individual NIMS pixels was weak, \citet{hansen-2004-Amorphouscrystallinea} did not present a global map of this feature, but instead reported averages of icy pixels from five large regions.  Other maps of \ce{H2O} spectral features using NIMS data include those by \citet{hibbitts-2000-DistributionsCO2,mccord-1998-Nonwatericeconstituents,stephan-2020-H2Oiceparticle}. In this work, we primarily use the 1.48 \mc{} water ice band depth map in \citet{hibbitts-2000-DistributionsCO2} as a point of reference, as the band depths are presented as a global map, allowing for a straightforward comparison between our data and the Callisto geologic map \citep{greeley-2000-Galileoviews}.
 
\emph{Interpretation of Regional Fresnel Peak Trends (Leading):} As shown in Fig.~\ref{jwst-fig:fresnel-maps}, the Fresnel peak band appears correlated with regional geology on Callisto's leading hemisphere in both the Valhalla-centered and Asgard-centered observations. The intensity is stronger in the large impact basins, Asgard and Valhalla (Fig.~\ref{jwst-fig:fresnel-maps}). This finding is consistent with their generally bright albedos and previous evaluations of Callisto's \ce{H2O} ice distribution, including in \citet{hibbitts-2000-DistributionsCO2}. For \edit1{these features}, the pixels with the \edit1{largest} Fresnel peaks appear to be contained to the center regions of the craters, with a drop off in strength toward the outer rims. This is mirrored in the albedo maps, which show the brightest material of these craters is retained in the impact centers. Supporting these JWST results are millimeter/submillimeter observations of Callisto using ALMA, which show the Valhalla impact region is co-located with a thermal cold spot discernible in model residuals spanning 3--0.87 mm \citep{camarca-2025-MultifrequencyGlobal,camarca-2023-ThermalProperties}. This cold spot may be explained if there is higher thermal inertia material present that suppresses daytime temperatures in the crater center, and/or lower emissivity materials, both of which are consistent with a higher abundance of fresh, exposed \ce{H2O} ice. Moreover, the dimensions of the JWST NIRSpec pixel are comparable to the ALMA beamsize for the Callisto observations (within a factor of $\sim$2.5), and we find the spatial extent of the largest Fresnel peaks in this area to be comparable to the size of the cold spot in the ALMA data (of order \textless \ang[exponent-mode=input]{;;0.3}).

The large-impact enhancement in the Fresnel peak intensity is accompanied by smaller geologic units with similar levels of enhancement. On the Asgard-centered observation, the other group of more intense Fresnel peaks located \edit1{in the southern hemisphere near the sub-observer longitude} may be associated with a pair of overlapping light plains units noted in the \citet{greeley-2000-Galileoviews} geologic map. This pixel group appears to extend close to \textminus\ang[exponent-mode=input]{60}N; however, interpreting this region further is complicated by the geologic map at this longitude being cut off close to \textminus\ang[exponent-mode=input]{40}N because of poor spatial resolution. On the Valhalla-centered observation, there are several pixel groups with elevated Fresnel intensity that are likely sampling smaller impact features. The Fresnel peak enhancement just north of Valhalla may be linked to \edit1{a relatively well-preserved geologic unit} containing two craters near \ang[exponent-mode=input]{38}N, \ang[exponent-mode=input]{65}W that lies on the boundary of the Valhalla inner and outer crater units \citep{greeley-2000-Galileoviews}. Southwest of the Valhalla center, between about 55 to \ang[exponent-mode=input]{90}W and 20 to \ang[exponent-mode=input]{57}S there are around 15 mapped crater units \citep{greeley-2000-Galileoviews} that may explain the Fresnel peak enhancement in this area in our maps. This particular section of Callisto's geologic map appears to bear a higher density of mapped young crater units than other similarly sized sections of the geologic map. The brightest of the craters in this spatial group, Agloolik (\ang[exponent-mode=input]{-47}N, \ang[exponent-mode=input]{82}W), appears to have partially saturated in the USGS albedo map. 

In the southern hemisphere on the Valhalla-centered observation, we recover elevated Fresnel band areas in the vicinity of Lofn. Lofn (\ang[exponent-mode=input]{-57}N, \ang[exponent-mode=input]{22}W) is one of Callisto's youngest large-scale features ($\sim$1.4--3.9 Gyr) and is geologically significant because its morphology points toward a fragmented parent body that impacted obliquely into a slushy or liquid zone \citep{greeley-2001-GeologyLofn}. In particular, the ejecta units inferred to be sourced from the greatest subsurface depths are interpreted to retain impact melt and, possibly, materials from a subsurface ocean \citep{greeley-2001-GeologyLofn}. Mapped units from Lofn partially obscure Heimdall to the southeast and Adlinda to the northwest. Heimdall is another large, bright impact feature, and Adlinda is possibly the oldest multi-ring structure on Callisto \citep{greeley-2001-GeologyLofn,wagner-1999-AgesIndividual}. Here, we note that the pixels we propose to be associated with Lofn/Heimdall are close to the edge of the disk, and therefore the spatial sampling of this region is poor. Consequently, it is difficult to infer the separate contributions of Lofn/Heimdall to the Fresnel peak. However, we infer that the contribution of Adlinda is probably not as important based on its lower albedo and less ice-rich spectra compared to Lofn (e.g., see Plate 2 in \citealt{greeley-2001-GeologyLofn}.) In the NIMS dataset, the Fresnel peak was extracted from the Lofn region as the representative high-latitude, sub-Jovian spectra \edit1{(\citealt{hansen-2004-Amorphouscrystallinea}, Fig.~8 therein)} and exhibits the same two-component structure in the same wavelength region as our extractions (Fig.~\ref{jwst-fig:fresnel-spectra}). Altogether, we find the Lofn/Heimdall region, along with the large multi-ring basins and smaller crater units described in the previous paragraph, are the primary terrains with elevated \threeum{} Fresnel band areas on the leading hemisphere. We now turn to describing a region with depleted \threeum{} Fresnel peaks on the leading hemisphere.

\emph{\edit1{Another region of interest:}}
\edit1{A portion of Callisto's surface} just east of Valhalla stands out for its weaker Fresnel band areas. This ``depleted'' region is fairly large compared to the spatial scale of individual groups of high-intensity Fresnel peaks, and may exhibit some amount of curvature around Valhalla. The dominant geologic unit in this region is a cratered plains unit, the dominant terrain on Callisto as readily observed in the \citet{greeley-2001-GeologyLofn} geologic map. As summarized by \citet{greeley-2001-GeologyLofn}, the crater plains terrain appears to be less ice-rich and possibly more silicate-rich than the centers of the large icy impacts \citep{carlson-1996-NearInfraredSpectroscopy,mccord-1997-OrganicsOther}. Although this context helps explain why this region bears depleted Fresnel peaks relative to the icy craters, more information is required to explain why this region is depleted compared to other swaths of leading hemisphere crater plains terrain. Our results do appear consistent with the albedo map, as this region appears darker than much of the cratered plains unit sampled by the Asgard-centered observation. 

Notably, this Fresnel-depleted region appears to overlap with a thermal anomaly reported in the ALMA data. In \citet{camarca-2023-ThermalProperties}, a region about $\sim$3 K warmer than model predictions (which account for albedo) in a 0.87 mm observation was noted in the region just southeast of Valhalla near \ang[exponent-mode=input]{-15}N, \ang[exponent-mode=input]{45}W; this warm terrain was also detected in the 1 and 3 mm data \citep{camarca-2025-MultifrequencyGlobal}. As described by \citet{camarca-2023-ThermalProperties}, this region may be consistent with locally low thermal inertia terrain, e.g., more texturized, loosely compacted regolith. Such regolith could be plausibly generated by preferential micrometeorite bombardment at low-latitudes on the leading hemisphere (a similar process was suggested to explain ALMA Ganymede observations;  \citealt{dekleer-2021-GanymedesSurface}). This micrometeorite hypothesis may be linked to some weak Fresnel peak measurements if the locally warm terrain has allowed for shallow ice to sublimate away. \edit1{Although impact gardening has been suggested as a mechanism of ice exposure, \cite{spencer-1984-Mobilitywater,spencer-1987-Thermalsegregation} determined the removal of ice via thermal segregation would dominate for Callisto’s lower latitudes.} A caveat to this explanation is that, although this region is observed on the leading hemisphere, it is not centered at Callisto's apex of motion (\ang[exponent-mode=input]{90}W). Altogether, this interpretation presents an alternative to the suggestion that impacts excavate subsurface \ce{H2O} ice, as was posited by \citet{bockelee-morvan-2024-Compositionthermal} to explain the Ganymede JWST Fresnel peak maps. Here, it is necessary to emphasize that the balance between the endogenic and exogenic origin of dark material on Callisto is a major outstanding question for this satellite, which motivates the acquisition of additional telescope observations for this region.

\edit1{\emph{No obvious ice migration}:} Another application of the Fresnel peak map is to evaluate an icy moon for evidence of the migration of ice crystals. As reported by \citet{bockelee-morvan-2024-Compositionthermal}, the strength of the Fresnel peak on Ganymede's trailing hemisphere appears to favor the morning limb. Such a distribution could be explained if the morning limb excess represents an accumulation of ice particles that have yet to be sublimated away in the daytime heat. In this work, we do not recover obvious evidence for a morning sublimation front on Callisto (Fig.~\ref{jwst-fig:fresnel-maps}). There is potential evidence for the afternoon limb on the trailing hemisphere bearing elevated Fresnel band areas. If this limb brightening was only controlled by geology, that unit would be located near \ang[exponent-mode=input]{210}W, and we do not recover  a similar limb brightening for this longitude at the morning limb of the Asgard-centered observation. \edit1{However, because the strength of Callisto's Fresnel peak is much weaker than those on Ganymede, pixels near the disk edge must be interpreted with greater caution. Therefore, we do not claim evidence of ice migration in this work.}

\emph{Ice Crystallinity on Callisto:} One of the \edit1{useful} diagnostics that can be derived from analysis of the Fresnel peak is whether the crystal is amorphous or crystalline. At \edit1{icy} Galilean satellite \edit1{surface} temperatures, the Fresnel peak is broad and weak for amorphous ice, and strong and structured for crystalline ice \citep{hansen-2004-Amorphouscrystallinea}. The identification of amorphous ice on the Galilean satellites is of interest given the dependence of ice crystallization on factors including temperature and the radiation environment. For example, \citet{jenniskens-1996-CrystallizationAmorphous} show that for a modest change in temperature from $\sim$105--125 K, the change in the crystallization timescale is a factor of $10^5$ faster. The laboratory work by \citet{jenniskens-1996-CrystallizationAmorphous} points to thermal recrystallization timescales for 150 K temperatures of order 10 seconds, with timescales under 1 minute for this same temperature including porosity-dependent measurements by \citet{mitchell-2017-Porosityeffects}. 
The paradigm established by analyses of the NIMS data is that radiation works quickly to produce amorphous ice on Europa, with some available on Ganymede, while thermal crystallization dominates at Callisto \citep{hansen-2004-Amorphouscrystallinea}. The presence of amorphous ice at Ganymede's north pole was recovered by \citet{bockelee-morvan-2024-Compositionthermal} using JWST NIRSpec. In the Ganymede data, the Fresnel peak is clearly broad with a lower band center reported at 3.05 \mc{}. Laboratory work by \citet{stephan-2020-H2Oiceparticle} demonstrates that the band center of the Fresnel peak can shift closer to 3.09 \mc{} in 150 K crystalline ice, which is broadly consistent with our band center measurements (Fig.~\ref{jwst-fig:fresnel-maps}).

\emph{Interpretation of Leading vs. Trailing:} Our finding that Fresnel peak band areas are smaller on the trailing hemisphere than on the leading hemisphere is consistent with the trend observed on Callisto's neighbor, Ganymede \citep{bockelee-morvan-2024-Compositionthermal}. This is apparent by comparing example leading/trailing hemisphere spaxels here (Fig.~\ref{jwst-fig:fresnel-spectra}), as well as in the disk-integrated leading/trailing spectra presented in Fig. 4 in \citet{cartwright-2024-RevealingCallistos} with Fig. 4 of \citet{bockelee-morvan-2024-Compositionthermal}. The contrast between Callisto's average leading/trailing hemisphere band areas as measured here ($\sim$1.4) is much lower than the contrast of $>$10 reported for Ganymede. As summarized in \citet{bockelee-morvan-2024-Compositionthermal}, the Fresnel peak may be strengthened on the leading hemisphere of Ganymede due to more ions sputtering fresh ice, as well as the effect of increased meteorite flux excavating the ice. \edit1{For Callisto, the distribution of sputtering ions and electrons likely affects surface processing, albeit to a lesser degree than its sibling Galilean moons (e.g., \citealt{cooper-2001-EnergeticIon,johnson-2004-Radiationeffects}). However, direct comparisons between Ganymede and Callisto are complicated by the former's internal magnetic field.} \edit1{Therefore, while Callisto's muted leading/trailing Fresnel peak asymmetry may partly reflect weakened energetic particle processes, further work is needed to clarify which aspects of the Jovian plasma environment are most relevant to the 3.1 \mc{} feature specifically.} 

\edit1{Notably, we recover a bullseye pattern, in which the Fresnel peak band area is smaller at the disk center and larger toward the edge, on Callisto's trailing hemisphere}. A similar trend is discernible in maps of the 1.48 \mc{} water ice band on Callisto in \citet{hibbitts-2000-DistributionsCO2}. Importantly, a trailing hemisphere Fresnel bullseye is not unique to Callisto, as the JWST map of this feature on Ganymede exhibits the same behavior \citep{bockelee-morvan-2024-Compositionthermal}. One possible explanation for the bullseye pattern is temperature: laboratory work by \citet{stephan-2021-VISNIRSWIR} demonstrates that the intensity of the Fresnel peak decreases with increasing temperature, which would predict a weaker equatorial Fresnel peak on Callisto. \edit1{However, if temperature were the primary factor varying the Fresnel peak, we would expect weak equatorial band areas in the Asgard- and Valhalla-centered observations, which we do not see}. Grain size is another factor to consider; the same laboratory data demonstrate that increasing the grain size of \ce{H2O} ice increases the Fresnel peak intensity across the 70--150 K range for grains of order 70--1300 \mc{} \citep{stephan-2021-VISNIRSWIR}. However, a detailed analysis of NIMS data by \citet{stephan-2020-H2Oiceparticle} using the 2 \mc{} and 1.5 \mc{} band features finds that Callisto's trailing hemisphere has an apparent bullseye pattern in grain size, with larger grains at the equator and smaller grains at the poles. Based on grain size alone, this would predict a larger Fresnel peak at the equator and a smaller one at the poles, which is opposite to our result, further indicating that additional processes must be at play. In the next section, we speculate on whether the large abundance of \ce{CO2} on Callisto's trailing hemisphere may be related.

\subsection{\texorpdfstring{Placing Our \ce{CO2} Results in Context}{Placing Our CO2 Results in Context}}
\emph{Solid-phase \ce{CO2}:}
We find that our mapping of \ce{CO2} on Callisto's Valhalla-centered hemisphere as observed with JWST NIRSpec is consistent with past observations of this volatile's distribution. Our results \edit1{agree} with the analyses of \edit1{\cite{mccord-1998-Nonwatericeconstituents} and \cite{hibbitts-2000-DistributionsCO2} who found that} \ce{CO2} band depths are weaker in Callisto's polar regions compared to lower latitudes. The average \ce{CO2} band depth measured for the Valhalla-centered observation of $\sim$19\% appears consistent with the \ce{CO2} map provided by \citet{hibbitts-2000-DistributionsCO2}, although the \emph{Galileo} map includes much finer spatial resolution observations that showcase elevated \ce{CO2} in smaller craters well below the resolution of a NIRSpec pixel. The inter-crater band depths in the \citet{hibbitts-2000-DistributionsCO2} map appear to be $\sim$15\%, while the dotted areas of elevated \ce{CO2} in small craters appears to be $\sim$30\%. The average \ce{CO2} band depth measured for this Valhalla observation is much lower than those measured on the trailing hemisphere using JWST data by \citet{cartwright-2024-RevealingCallistos}, again consistent with past analyses \citep{hibbitts-2000-DistributionsCO2,mccord-1998-Nonwatericeconstituents}. 

\begin{figure}
\centering
\includegraphics[width=1\textwidth]{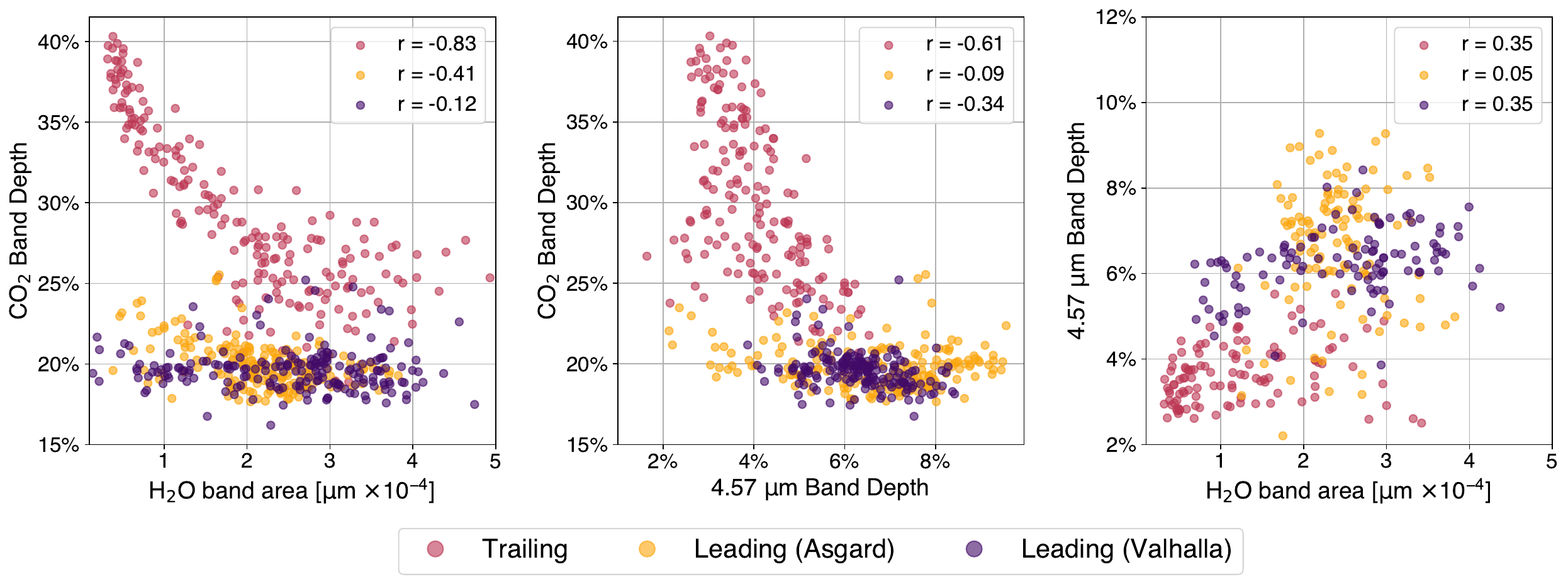}
\caption{\edit1{Correlation diagrams for the strength of three spectral features surveyed on Callisto in this work, namely the 4.25 \mc{} \ce{CO2} absorption, the 3.1 \mc{} \ce{H2O} ice Fresnel peak, and the 4.57 \mc{} absorption band.} The presented r values are the Spearman rank-order correlation coefficients, \edit1{which} can vary between -1 and +1, where 0 implies no correlation.}
\label{jwst-fig:co2-corr}
\end{figure}

\begin{figure}
    \centering 
    \includegraphics[width=0.3\textwidth]{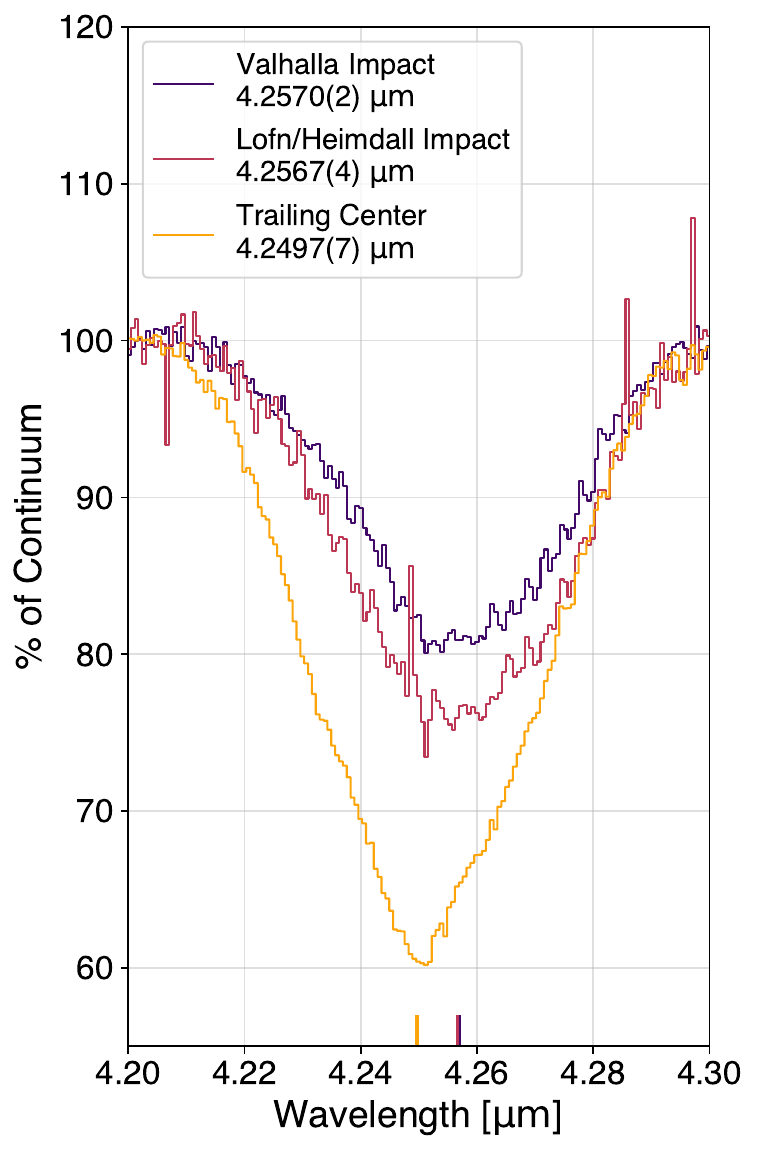}
    \caption{Sample continuum-divided \ce{CO2} absorption features on Callisto. Sampled regions include a spaxel from near the center of the Valhalla impact basin, a spaxel from the \ce{CO2} enrichment near the Lofn/Heimdall impacts in the southern hemisphere, and a spaxel from the center of the trailing hemisphere. The vertical lines at the bottom of the plot indicate the best-fit band center for each spaxel. \edit1{The uncertainties in the band centers are given in parentheses.} }
    \label{jwst-fig:co2-spectra}
\end{figure}

Moreover, we find that our recovery of an enrichment of \ce{CO2} in the vicinity of the large Lofn/Heimdall impacts agrees well with \emph{Galileo} results. Our measured band depths in the range of 25\% are comparable to those measured with NIMS, although there are several regions below the NIRSpec resolution in the \citet{hibbitts-2000-DistributionsCO2} map which exhibit band depths upwards of 40\%. In the \citet{hibbitts-2002-CO2richimpact} dedicated analysis of \ce{CO2}-rich craters on Callisto, Lofn is not specifically analyzed, nor is the origin of \ce{CO2} in Lofn specifically considered by \citet{greeley-2001-GeologyLofn}. \edit1{As shown in Fig.~\ref{jwst-fig:fresnel-maps}, both the center of Valhalla and Lofn also bear elevated Fresnel peak areas, which means these large impacts are relatively rich in icy host material for \ce{CO2}. Correspondingly, we show in Fig.~\ref{jwst-fig:co2-spectra} that the extracted \ce{CO2} spectra of these impacts are both red-shifted in comparison to the radiolytically generated trailing hemisphere \ce{CO2}. } As discussed earlier, Lofn may retain \ce{H2O} ice melt possibly derived from a subsurface liquid/slushy zone. Therefore, we evaluate the NIMS dataset in combination with these new JWST results to evaluate the origin of \ce{CO2} in the vicinity of Lofn.

 First, we consider the evidence for \ce{CO2} derived from a subsurface liquid/slushy zone. \edit1{It is important to emphasize we are not exploring recent delivery, as the estimated minimum age of Lofn is $\sim$1.4 Gyr \citep{wagner-1999-AgesIndividual,wagner-1998-TimeStratigraphyCrater}.} Here, we evaluate if the \ce{CO2} in Lofn/Heimdall appears connected to the water-rich ejecta deposits. Based on the high-resolution NIMS dataset, it is not obvious this is the case. There are several units of icy impact ejecta described in the \citet{greeley-2001-GeologyLofn} Lofn geologic map, with strong \ce{H2O} signatures in the crater center. By contrast, in the NIMS map, \ce{CO2} is distributed very broadly across the entire Lofn unit with no apparent correlation with the excavated \ce{H2O} ice or \edit1{an increase in band depth at the crater center}. As described by \citet{hibbitts-2002-CO2richimpact}, \ce{CO2} rich craters on Callisto may follow a depletion pattern whereby the ice is first depleted from the ejecta blanket, then the crater rays, and lastly the crater itself. A subsurface liquid origin for the \ce{CO2} in Lofn is also complicated by the fact the much older, adjacent Adlinda multi-ring basin is similarly enriched in \ce{CO2} in the NIMS map. As such, in light of the present data, a subsurface origin for the majority of Lofn's \ce{CO2} seems unlikely, \edit1{thereby requiring a different explanation for its origin.} It has been demonstrated that thermal properties of large impact features on Ganymede and Callisto are consistent with the presence of locally high thermal inertia materials (\citealt{dekleer-2021-GanymedesSurface,camarca-2023-ThermalProperties,camarca-2025-MultifrequencyGlobal}). The Lofn/Heimdall/Adlinda region is observed to be colder in model residuals at 3--0.87 mm than other \edit1{terrains of similarly high latitude, suggesting that \ce{CO2} may be preferentially cold trapped in these areas}. 

 \emph{\ce{CO2} and exposed \ce{H2O} ice on the trailing hemisphere: } 
 Earlier in the discussion section, we highlighted that the trailing hemisphere Fresnel peak distribution appears to be an ``inverted'' bullseye pattern, with depletion at the disk center (Fig.~\ref{jwst-fig:fresnel-maps}). Here, we compare this pattern to Callisto's well-known \ce{CO2} bullseye pattern, which features an enrichment at the disk center. In Fig.~\ref{jwst-fig:co2-corr}, we show that the strength of the Fresnel peak band area is anti-correlated with the band depth of the \ce{CO2} feature. If we interpret this anti-correlation as a tradeoff between the relative abundance of \ce{CO2} and \ce{H2O} ice on Callisto's trailing hemisphere, there are several possible explanations. First, the production of the \ce{CO2} on Callisto's trailing hemisphere is radiolytic, and may proceed via chemistry between dark, C-rich material and \ce{H2O} ice, a process supported by laboratory investigations (e.g., \citealt{raut-2012-RADIATIONSYNTHESIS,mennella-2004-FormationCO,hand-2012-Laboratoryspectroscopic}). Therefore, the areas of high radiolytic \ce{CO2} production may have depleted a region of feeder \ce{H2O} molecules. However, \citet{moore-2004-Callisto} suggest that irradiated \ce{H2O} ice is not sufficient to explain the \ce{CO2} abundance as it would require the \ce{H2O} ice to migrate into the dark, non-ice material before escaping to space. 
 Results from \emph{Galileo} demonstrate that \ce{H2O} ice and non-ice material exist on scales below a few km (the best NIMS resolution), however higher-resolution Solid State Imager (SSI) images show light/dark patches at the scale of tens of meters, and areal mixing might even occur at cm levels \citep{moore-2004-Callisto}. \edit1{Nevertheless, while radiolytic generation of \ce{CO2} from \ce{H2O} could be a relevant factor in understanding Callisto’s trailing hemisphere Fresnel peaks, the same is not obvious for Ganymede. JWST observations of Ganymede reveal that this moon also possesses a trailing hemisphere bullseye pattern \citep{trumbo-2023-Hydrogenperoxide,bockelee-morvan-2024-Compositionthermal}, but  the 3.1 \mc{} peak areas are weakly correlated with 4.25 \mc{} \ce{CO2} band depths \citep{bockelee-morvan-2024-Compositionthermal} As such, a mechanism common to Ganymede and Callisto may be required to explain the nature of trailing hemisphere ice in the Galilean system. A direct comparison with Europa will also be useful once the existing JWST analyses of \ce{CO2} \citep{trumbo-2023-distributionCO2,villanueva-2023-EndogenousCO} and the Fresnel peak \citep{cartwright-2025-europa-fresnel} are complemented with results from the trailing hemisphere. Clearly, the story of radiolytic \ce{CO2} generation and its possible links to interpreting surface ice measurements is further complicated across temperature, grain size, and areal exposure constraints. Future work to tether surface chemistry with the proper sources and sinks of \ce{CO2} and \ce{H2O} on Callisto’s trailing hemisphere, in the context of the local plasma environment, would be useful in understanding this puzzling moon.  }

\begin{figure}
\centering
\includegraphics[width=1.0\textwidth]{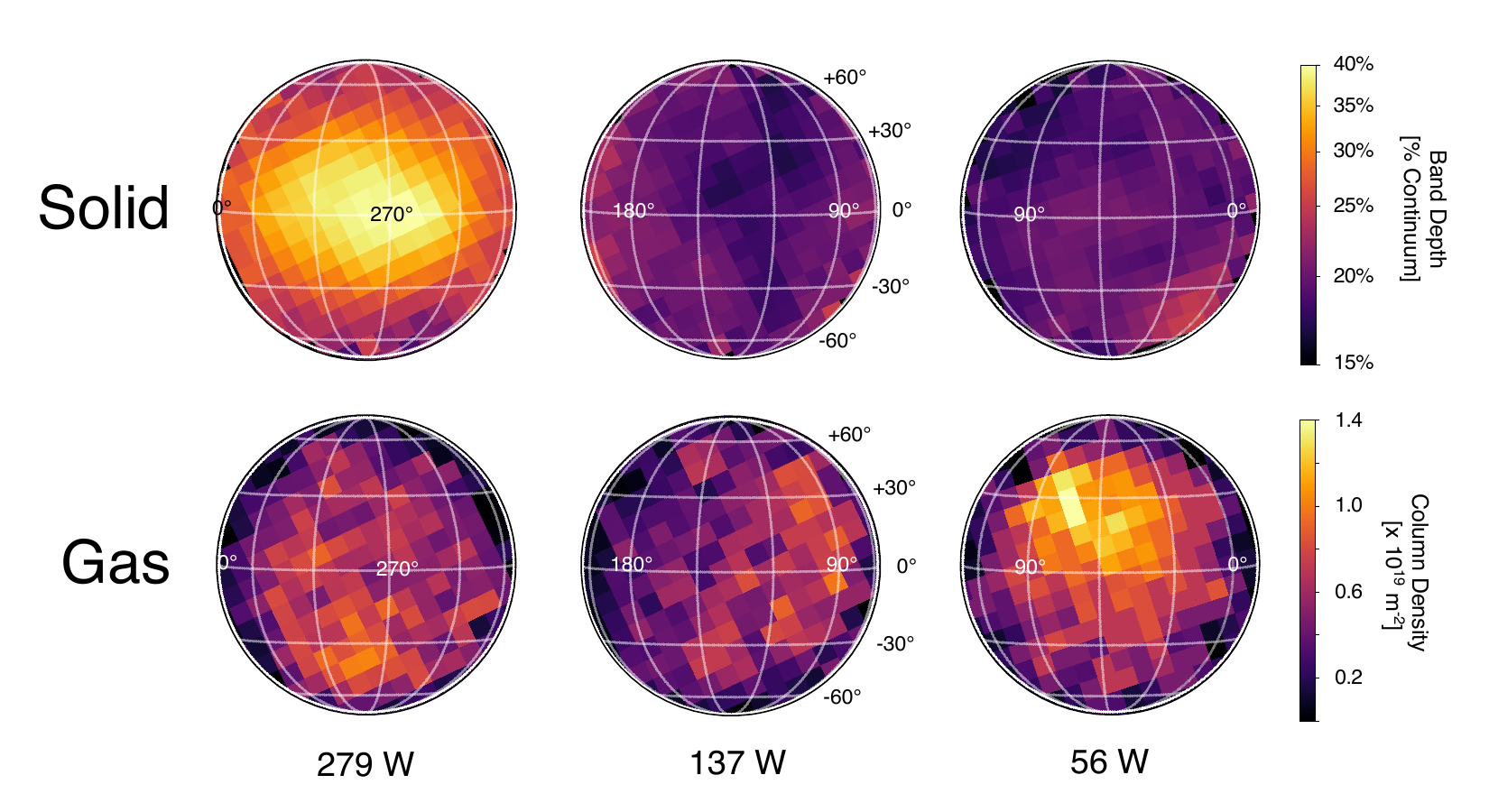} 
\caption{The distribution of solid-phase and gaseous \ce{CO2} on Callisto. Callisto north is up. Data from the first two columns were originally presented in \cite{cartwright-2024-RevealingCallistos}, data in third column from this work.  }
\label{jwst-fig:gas-and-solid-co2}
\end{figure}

\emph{Gas-phase \ce{CO2}:} 
The original detection of \ce{CO2} gas at Callisto was reported by \cite{carlson-1999-TenuousCarbon}, who used an off-limb scan from \emph{Galileo} during its C10 flyby. \cite{carlson-1999-TenuousCarbon} inferred a column density of \SI{0.8e19}{m^{-2}}, and proposed that the volatility and mobility of \ce{CO2} indicates it should be globally present on Callisto. These major attributes of the atmosphere were recently confirmed using new JWST observations. As determined by \cite{cartwright-2024-RevealingCallistos}, \ce{CO2} gas is present across both Callisto’s leading and trailing hemispheres, and the retrieved column densities 0.4--\SI{1.0e19}{m^{-2}},  are similar to those reported by Galileo. The fact that the column densities in the JWST data were substantially weaker off the disk supports \cite{carlson-1999-TenuousCarbon}'s conclusion that \ce{CO2} gas is retained close to the surface (\textless40 km), \edit1{an unsurprising result given Callisto's large mass.  Our work adds a third JWST detection of \ce{CO2} gas on Callisto, now on the Valhalla-centered hemisphere} (Fig.~\ref{jwst-fig:co2-gas},~\ref{jwst-fig:gas-and-solid-co2}).  Our reported column densities range from $\sim$0.4--\SI{1.4e19}{m^{-2}}, and are comparable to the previous \cite{cartwright-2024-RevealingCallistos} values for the trailing and Asgard-centered hemisphere. \edit1{We note that, given the size of JWST's spatial footprint at the distance of Callisto, we do not attempt an exact measure of the scale height in this work}. \edit1{To put the retrieved \ce{CO2} column densities in context, the inferred abundances of two other main species in Callisto’s atmosphere are \ce{O2} $\sim$ \SI{4e19}{m^{-2}} and \ce{H2}$\sim$\SI{e19}{m^{-2}}. The value for \ce{O2} was inferred from excited atomic oxygen emissions \citep{cunningham-2015-DetectionCallistos,dekleer-2023-OpticalAurorae}, and the constraint for \ce{H2}$\sim$ \SI{e19}{m^{-2}} was modeled by  \citep{carberrymogan-2022-h2} based on coronal H emission observed by \cite{roth-2017-Detectionhydrogen}. Altogether, Callisto’s \ce{CO2} is seemingly not the dominant atmospheric species, but is at most about an order of magnitude less abundant than \ce{O2} (and potentially only a few times less). \edit1{Being the heaviest among the main species (\ce{O2}, \ce{H2}), \ce{CO2} may be most relevant} \edit1{near the surface} and as collisional partner for other molecules. }

Beyond these global results, the variation of \ce{CO2} gas across the disk in context with the solid-phase \ce{CO2} reveals a spatial pattern for which the origin is not yet clear. As shown in Fig.~\ref{jwst-fig:gas-and-solid-co2} and first highlighted by \cite{cartwright-2024-RevealingCallistos}, \ce{CO2} gas across Callisto is patchy. Across the observations, the position of diffuse gas patches does not correspond to the location of deepest 4.25 \mc{} \ce{CO2} band depths, nor does there seem to be a correlation with the warmest regions, on the afternoon side of the subsolar point \citep{cartwright-2024-RevealingCallistos}. A study of \ce{CO2} gas on Ganymede yielded a similar general result, in that reported peak \ce{CO2} column densities do not overlap with the warm, equatorial regions nor with the largest \ce{CO2} band depths \citep{bockelee-morvan-2024-patchyCO2}. For Ganymede, the region of highest \ce{CO2} column densities is congregated in the north polar region on the leading hemisphere, while for Callisto we find that broadly the terrain around Valhalla is co-located with Callisto's elevated column densities \edit1{(Fig.~\ref{jwst-fig:gas-and-solid-co2})}.

If the Valhalla region is a source of gaseous \ce{CO2} on Callisto's leading hemisphere, an explanation is needed to explain why the Lofn/Heimdall region is seemingly insignificant in terms of \ce{CO2} gas despite being the primary solid \ce{CO2} reservoir on this face of the moon. One explanation might be that the Lofn/Heimdall region is at a geographic disadvantage with regard to participating in the processes expected to drive the transfer of \ce{CO2} to the atmosphere from the surface--\edit1{such as sublimation, which is more relevant at equatorial latitudes}. However, this does not explain why peak \ce{CO2} gas at Callisto is observed at high ($\gtrsim$\ang[exponent-mode=input]{30}) latitudes on both hemispheres. The role that Callisto's thermal properties play in organizing the \ce{CO2} gas across the surface remains unclear in light of this new Valhalla-centered observation. \cite{camarca-2025-MultifrequencyGlobal} suggested the trailing hemisphere \ce{CO2} gas peak in \citet{cartwright-2024-RevealingCallistos} might be related to a cold spot observed in the 3 mm trailing hemisphere residuals. Alternatively, \citet{cartwright-2024-RevealingCallistos} suggested that the warm, sub-Valhalla region detected in the ALMA residuals might promote sublimation of \ce{CO2} in those regions. The release mechanism of \ce{CO2} is likely complicated, much like the \ce{CO2} gas on its sister moon, Ganymede \citep{bockelee-morvan-2024-patchyCO2}. Of note, these observations are only ``snapshots'' of the atmosphere---if \ce{CO2} gas is mobile across Callisto, then our ability to diagnose \edit1{whether the gas is connected to specific terrain or not} is limited.  

\subsection{The \fourmic{} Feature and Organics on Callisto}

The existence of CN-bearing material on Callisto was proposed over 30 years ago at the discovery of the \fourmic{} feature in \emph{Galileo} data \citep{mccord-1997-OrganicsOther}. Subsequent observations \citep{cartwright-2020-EvidenceSulfurbearing}, including those using JWST \citep{cartwright-2024-RevealingCallistos}, continue to support CN as a candidate for the \fourmic{} absorption. \edit1{If this assignment holds, this absorption may offer a rare glimpse into nitrogen-based chemistry in the Galilean moon system, and may be comparable to the 4.62 \mc{} band detected in  astrophysical ices.} First observed in a protostar by \cite{soifer-1979-48micron}, the 4.62 \mc{} feature was later attributed to the OCN\textsuperscript{$-$} ion by \cite{grim-1987-Ionsgrain}. \edit1{From the discovery of Callisto's \fourmic{} band, it was suggested that the responsible compound may have a corresponding feature in interstellar ice spectra \citep{mccord-1997-OrganicsOther}.} Many decades of laboratory work confirm the relationship of OCN to the 4.62 \textmu{m} feature \citep{dhendecourt-1986-Timedependentchemistry,demyk-1998-Laboratoryidentification,hudson-2001-FormationCyanate,gerakines-2025-Establishingaccurate}. Today, the catalog of OCN detections includes young stellar objects \citep{vanbroekhuizen-2005-35$mathsfmu$m}, and molecular clouds \citep{mcclure-2023-IceAge}, and even on trans-Neptunian objects \citep{cryan-2025-NitrogenbearingSpecies}. 

By contrast, detections of bands near \fourmic{} on icy moons are not common. For example, in the Saturnian system, a narrow band near 4.53--4.55 \mc{} was reported in Cassini VIMS observations of Phoebe \citep{coradini-2008-Identificationspectrala}, however it was not recovered in recent JWST observations of this object \citep{belyakov-2025-SaturnianIrregular}. A weak \fourmic{} absorption \edit1{was reported} in Ganymede NIMS observations \citep{mccord-1997-OrganicsOther}, however a subsequent detection was not reported in JWST observations \citep{bockelee-morvan-2024-Compositionthermal} nor in the Juno close fly-by observations using the \edit1{Jovian Infrared Auroral Mapper instrument} \citep{tosi-2024-CharacterizationSurfaces}. The widespread detection of the \fourmic{} feature on Callisto is clearly unique, and the updated JWST results here offer useful context in developing a working hypothesis for its origin.

Previous work has highlighted the potential role of Jovian irregular satellites in delivering CN-bearing dust to Callisto \citep{cartwright-2024-RevealingCallistos,cartwright-2020-EvidenceSulfurbearing}. There are a few lines of spatial evidence that may be consistent with this hypothesis.  First, the primary global attribute of the strength of Callisto's \fourmic{} is deeper band depths on the leading hemisphere \citep{cartwright-2020-EvidenceSulfurbearing,cartwright-2024-RevealingCallistos}. Irregular satellite dust is generally expected to primarily fall on Callisto's leading hemisphere \citep{bottke-2013-Blackrain,chen-2024-Lifedust}, which is consistent with a proposed relationship between the \fourmic{} feature and the irregular satellite population. Second, \cite{davis-2025-SpectroscopicMapping} proposes that the muted \fourmic{} band depths observed in Asgard and at the edge of Valhalla impact basins in \cite{cartwright-2024-RevealingCallistos} might even be evidence of a ``timestamp" on the dust delivery. Under this paradigm, the rapid accretion of Jovian irregular satellite dust in the early stages of Callisto's \edit1{surface history} \citep{bottke-2013-Blackrain} was locally interrupted by the large impact basins. This work, which includes a more favorable viewing geometry of Valhalla, suggests that both this basin and Asgard could be depleted in the \fourmic{} absorption (Fig.~\ref{jwst-fig:456-results}). 

\edit1{While the distribution of Callisto's \fourmic{} band might be consistent with delivery of CN-bearing dust from the Jovian irregular satellites, this idea can be further evaluated with new compositional data}. Using JWST \edit1{NIRSpec}, \citet{sharkey-2025-JWSTReveals} procured 0.7--5.2 \mc{} reflectance spectra of eight Jovian irregulars, including three members from the dominant satellite family by mass, the Himalia family. \citet{sharkey-2025-JWSTReveals} report the surveyed Jovian irregulars do \emph{not} possess a \fourmic{} absorption feature. This result suggests that the observed Jovian irregular satellite families are not delivering CN-bearing dust ``as is'' to Callisto. However, \cite{sharkey-2025-JWSTReveals} found that some irregular satellites, including Himalia, do possess absorption features at 2.7 and 3.05 \mc{} consistent with the presence of ammoniated phyllosilicates. As such, \cite{sharkey-2025-JWSTReveals} suggest that perhaps the Jovian irregulars are pollinating Callisto with N-bearing minerals that react with C-rich material on its surface to yield CN-bearing products. It is useful to consider what material is providing the C, and under what conditions.  In our study, we find the only link between the \fourmic{} feature and \ce{CO2} is a modest (r = --0.61), inverse relationship on the trailing hemisphere \edit1{(Fig.~\ref{jwst-fig:co2-corr})}. Because the \ce{CO2} on Callisto's trailing hemisphere may be radiolytically generated, our result suggests the trailing hemisphere particle environment might be hostile to certain reactions between C-rich material and N-bearing material.

Clearly, much work remains to evaluate the origin of the \fourmic{} band on Callisto. \edit1{} Better hypothesis testing will hinge on laboratory research to study C and N chemistry using temperature and radiation conditions appropriate for an icy moon like Callisto.

\section{Conclusions}\label{jwst-conclusion}

The distribution of surface volatiles on Callisto's leading hemisphere was examined using the NIRSpec instrument onboard JWST. Here, we presented 1) the first global, high-resolution, high-sensitivity map of the water ice  Fresnel reflection peak at \threeum{} across Callisto's entire surface, and 2) a new map of spectral features at 4.25 \mc{} (\ce{CO2}) and \fourmic{} on a Valhalla-centered viewing geometry. For \ce{CO2}, we mapped both the solid-phase distribution as well as the gaseous-phase column densities. From these data products, we concluded the following:

--- The strength of Callisto's \threeum{} Fresnel peak tracks primarily with geology (i.e., large impacts) on the leading hemisphere. Valhalla, Asgard, and Lofn/Heimdall are all terrains with increased Fresnel band areas. This finding is consistent with their high albedoes and previously measured \ce{H2O} ice content from the NIMS instrument onboard \emph{Galileo} \citep{moore-2004-Callisto,hibbitts-2000-DistributionsCO2}. On the trailing hemisphere, the Fresnel peak exhibits an ``inverted bullseye'' pattern, with the band areas being weaker toward the more equatorial latitudes.

--- The greatest abundance of \ce{CO2} on Callisto's leading hemisphere is co-located with the Lofn/Heimdall impact basins. The shape and band center of the \ce{CO2} spectral feature in this region is comparable to that in the Valhalla impact basin, suggesting the two impacts host their \ce{CO2} in similar (icy) material. Lofn/Heimdall may host the largest share of non-radiolytic \ce{CO2} on Callisto. It is possible that \ce{CO2} in the Lofn/Heimdall region has remained enriched relative to the rest of the leading hemisphere given its location at high latitudes, where mechanisms that would otherwise release volatiles from the surface—such as solar insolation—are weakened. The thermal properties of the region as inferred from ALMA observations \citep{camarca-2023-ThermalProperties,camarca-2025-MultifrequencyGlobal} may further reinforce the retention of \ce{CO2} in this area through lowering peak surface temperature. Fortunately, Lofn and Heimdall have been identified as ``Regions of Interest'' on Callisto for \edit1{ESA's Juice (Jupiter Icy Moons Explorer)} mission and its JANUS (Jovis Amorum ac Natorum Undique Scrutator) camera will image this region at $\sim$3--4 km per pixel, and the visible and infrared Moons And Jupiter Imaging Spectrometer (MAJIS) will achieve 30 m/pixel resolution data \citep{stephan-2021-Regionsinterest}.

--- On the trailing hemisphere, we identify that the band depth of the solid-phase \ce{CO2} is anti-correlated with the strength of the Fresnel peak band area. The regions of highest \ce{CO2} absorption at the low latitudes centered on the trailing hemisphere are the regions of lowest Fresnel peak strength, and both molecules exhibit a characteristic ``bullseye'' pattern. This anti-correlation could be explained if the radiolytic processes on Callisto's trailing hemisphere regolith are depleting the region of its exposed \ce{H2O} ice, an ingredient in the generation of \ce{CO2}. Alternatively, \ce{H2O} ice could be removed from low-latitude regions by sublimation promoted by the highly volatile nature of \ce{CO2}, or perhaps is otherwise altered by charged particles that interact with the trailing hemisphere.

--- We detect a patchy \ce{CO2} atmosphere with peak column densities $\sim$\SI{1.4e19}{m^{-2}} on the Valhalla-centered observation. We find that the peak column densities of the gaseous \ce{CO2} are offset from the region of strongest solid-phase \ce{CO2} absorption. A similar offset was noted for  Callisto's trailing hemisphere \citep{cartwright-2024-RevealingCallistos}. Future modeling efforts as well as telescopic observations are needed to understand the mechanisms driving \ce{CO2} release from the surfaces of icy moons like Callisto.

--- We find that the band depth of the \fourmic{} feature is stronger on Callisto's leading hemisphere, including with the new Valhalla-centered observation, than on the trailing hemisphere. The material responsible for the \fourmic{} absorption, which may be a CN-bearing molecule (e.g., \citealt{cartwright-2024-RevealingCallistos}), and the recent discovery of ammoniated silicates on the Jovian irregular satellites \citep{sharkey-2025-JWSTReveals} offers a potential source of Nitrogen for Callisto's surface. Additional labwork using temperature and radiation conditions appropriate for Callisto is required to unravel the origin of its \fourmic{} absorption.

\section*{Acknowledgments}
This research is based on observations made with the NASA/ESA James Webb Space Telescope obtained from the Space Telescope Science Institute, which is operated by the Association of Universities for Research in Astronomy, Inc., under NASA contract NAS 5–26555. These observations are associated with program 2060 (JWST-GO-02060.001-A). M.C. acknowledges support from the National Science Foundation through a Graduate Research Fellowship, grant No. DGE-1745301. K.d.K. and M.C. acknowledge support from the National Science Foundation through grant No. 2308280. R.J.C. acknowledges support from Space Telescope Science Institute grant JWST-GO-02060.009-A. L.R. appreciates support from the Swedish National Space Agency through grant 2024-00112. K.P.H. was supported by the Jet Propulsion Laboratory, California Institute of Technology, under a contract with the National Aeronautics and Space Administration (80NM0018D0004). We also thank the reviewers for comments that improved the quality of this article. 

M.C. thanks Ryleigh Davis and Matthew Belyakov for helpful conversations about Callisto and JWST.

\bibliography{references.bib}
\bibliographystyle{aasjournal}
\end{document}